\documentclass[prl,twocolumn,preprintnumbers,amsmath,amssymb]{revtex4}%preprint,showpacs,twocolumn
\DeclareMathOperator{\tint}{\textstyle\int}
\usepackage{bm,graphics}
\DeclareSymbolFontAlphabet{\mathbfcal}{boldsymbols}
\begin{document}
\title{On the linear response and scattering of an interacting molecule-metal system}
%:\\Coupling of molecular and plasmonic resonances
\author{David J. Masiello}
\email{masiello@chem.northwestern.edu}
\author{George C. Schatz}
\email{schatz@chem.northwestern.edu}
\affiliation{Department of Chemistry, Northwestern University, Evanston, Illinois 60208-3113, USA}
\date{\today} 
\begin{abstract}
A many-body Green's function approach to the microscopic theory of plasmon-enhanced spectroscopy is presented within the context of localized surface-plasmon resonance spectroscopy and applied to investigate the coupling between quantum-molecular and classical-plasmonic resonances in monolayer-coated silver nanoparticles. Electronic propagators or Green's functions, accounting for the repeated polarization interaction between a single molecule and its image in a nearby nanoscale metal, are explicitly computed and used to construct the linear-response properties of the combined molecule-metal system to an external electromagnetic perturbation. Shifting and finite lifetime of states appear rigorously and automatically within our approach and reveal an intricate coupling between molecule and metal not fully described by previous theories. Self-consistent incorporation of this quantum-molecular response into the continuum-electromagnetic scattering of the molecule-metal target is exploited to compute the localized surface-plasmon resonance wavelength shift with respect to the bare metal from first principles.
\end{abstract}
%\pacs{33.20.Fb, 71.10. w, 31.15.xp}
\maketitle

Ever increasing experimental interest in a variety of plasmon-enhanced molecular spectroscopies has provided impetus for the development of a corresponding assortment of theoretical descriptions of these phenomena \cite{Corni2001a,Jensen06,Johansson2005,Neuhauser07,Masiello08a,AG09}. Linear molecular spectroscopies such as Raman and fluorescence have found their plasmon-enhanced analogs in surface-enhanced Raman scattering (SERS) and surface-enhanced fluorescence; both are now routinely realized in the extreme limit of single-molecule detection; see, e.g., Refs. \cite{Ginger07,Masiello08b,Feldmann08,Novotny06}. And plasmon-enhanced versions of $n$-wave mixing and hyper-Raman scattering \cite{Kelley06a,Chou09}, as well as other nonlinear spectroscopies, are rapidly being explored as ultrasensitive probes of molecular structure complementary to those linear. Conversely, spectroscopy of the plasmon itself, either localized to metal particles or propagating across metal surfaces \cite{kall09}, and embedded within a potentially resonant molecular environment \cite{RVD06,Zhao07} forms the basis for yet another promising direction of related research. Outside of the laboratory, the challenge remains to develop theories that are rich enough to describe the basic physics relevant to each, yet are simultaneously practical enough to implement numerically for real systems of current experimental interest.
%!!!!!!!!!!!!!!!!!!!!!!!!!!!!!!!!!!!!!!!!!!!!!!!!!!!!!!!!!!!
%!!!!!!!!!!!!!!!!!!!!!!!!!!!!!!!!!!!!!!!!!!!!!!!!!!!!!!!!!!!
\begin{figure}[t]
\begin{center}
\rotatebox{0}{\resizebox{!}{4.5cm}{\includegraphics{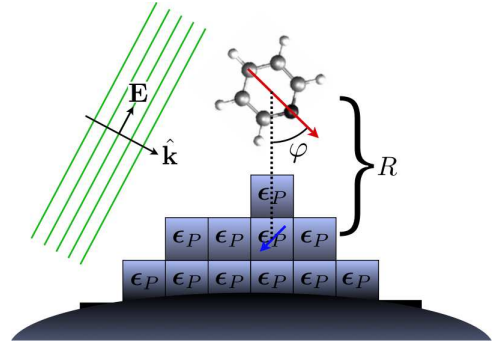}}}
\caption{\label{py} Polarization effects between a quantum-mechanical molecule and a classical nanoscale metal are built into the interacting one-body molecular Green's function $\cal G$ through the electron-plasmon potential $\hat V_P.$ Truncation of the associated self energy $\Sigma^\bigstar\approx\Sigma^\bigstar_{(2)}[{\bm\alpha}_P({\bm\epsilon}_P)]$ at second order in perturbation theory, followed by a partial resummation of the Dyson series for ${\cal G},$ allows the molecule to repeatedly polarize the metal's conduction electrons to all orders in perturbation theory. Underlying this interaction is the metallic polarizability ${\bm\alpha}_P(\omega),$ which is connected to the macroscopic permittivity ${\bm\epsilon}_P({\bf k}={\bf 0},\omega)$ of the metal through a particular model of dielectric response.}
\end{center}
\end{figure}
%!!!!!!!!!!!!!!!!!!!!!!!!!!!!!!!!!!!!!!!!!!!!!!!!!!!!!!!!!!!
%!!!!!!!!!!!!!!!!!!!!!!!!!!!!!!!!!!!!!!!!!!!!!!!!!!!!!!!!!!!

Here we present the first numerical implementation of a new and general many-body Green's function formalism that is capable of describing a variety of plasmon-enhanced linear spectroscopies and apply it to model recent localized surface-plasmon resonance (LSPR) spectroscopy and molecular sensing experiments. LSPR spectroscopy is highly sensitive to the small refractive-index changes of the environment surrounding nanoscale metal particles and has been reported to detect the presence of as few as tens to hundreds of nearby molecules \cite{Hook09,Chilkoti09}. This sensitivity arises from the resonant coupling of incident light to the collective oscillation of conduction electrons in the metal modified by the presence of the surrounding medium, which may have its own distinct electronic resonances. For metal spheres small in comparison to the wavelength of light, this resonance condition would be met at those frequencies $\omega_*$ in which
\begin{equation}
\label{sphere}
\textrm{Re}\{\epsilon_P(\omega)/\epsilon_M(\omega)\}=-2
\end{equation}
is satisfied, where $\epsilon_P$ and $\epsilon_M$ are the dielectric functions of the metal particle and molecular environment. Indeed, even in this limit, Eq. (\ref{sphere}) is difficult to solve when both the metal and environment exhibit resonant behavior in the same spectral regime.

We calibrate our first numerical description of LSPR spectroscopy through an approximate modeling of a series of recent Van Duyne experiments \cite{RVD06,Zhao07} probing the influence of electronically-resonant molecules upon the LSPR of bare silver nanotriangles. Emphasis is placed not upon the exact modeling of any one of these experiments but rather upon understanding behavior generic to all. To this end, we choose to replace the reported [2,3,7,8,12,13,17,18-octakis(propyl)porphyrazinato]magnesium(II), iron(II) tris-2,2'-bipyridine, and rhodamine 6G with the simpler pyridine molecule, treated fully quantum mechanically, and to approximate the silver nanotriangles by classical silver nanospheroids, which also have tunable aspect ratios and resonances; as in the experiments, this tunability will allow us to tune the LSPR into or detune away from one or more of the molecular resonances. To our knowledge there are no previous first-principles models of plasmon-enhanced spectroscopy that have investigated the self-consistent coupling between molecular and true plasmonic resonances within the context of LSPR spectroscopy for real systems like this. And it is a main point of this paper to report the achievement of such a computation.

By first principles we mean a theoretical description of the molecule-metal system in which the mutual coupling between molecule and metal and its effect on the molecular-electronic structure and electromagnetic-scattering properties of the combined target is computed without varying parameters to fit experiment, while the underlying response of the metal is completely governed by the phenomenological dielectric data for bulk metals; see, e.g., Ref. \cite{palik}. Other noteworthy theoretical approaches (some of which are first principles) have been applied to a variety of plasmon-enhanced processes such surface-enhanced Raman \cite{Corni2001a,Jensen06,Masiello08a} and fluorescence spectroscopy \cite{Johansson2005} as well as the transfer of energy through arrays of metal nanoscale particles \cite{Neuhauser07}. For example, Corni and Tomasi extend the polarizable-continuum model of molecular solvation to treat the interaction of a general molecular-electronic system with a nearby classical polarizable metal particle and compute Raman enhancement factors as a function of excitation energy \cite{Corni2001a}. While Jensen et al. compute SERS spectra from a variety of molecules adsorbed onto small metal clusters, intended to mimic nanoscale metal particles, with modern computational-chemistry methods, but are forced to empirically parametrize the polarization interaction between the molecule and the remainder of the metal \cite{Jensen06}. Johansson et al. describe both SERS and fluorescence of a two-state molecular-electronic system placed in the junction between two metal spheres within a density-matrix formalism that phenomenologically includes both radiative and nonradiative relaxation mechanisms \cite{Johansson2005}. And, within the context of energy transfer, Neuhauser and Lopata combine a finite-difference time-domain electrodynamics description of the metal particle and external field with a random-phase approximation of a two-state molecular-electronic system's explicitly time-dependent dynamics, together with a phenomenological damping of the molecular excitations due to the presence of the metal \cite{Neuhauser07}. Lastly, Masiello and Schatz rigorously combine modern molecular electronic-structure theory with a continuum-electrodynamics description of the metal, including the self-consistent coupling between the two, to enable practical computation of SERS in real systems \cite{Masiello08a}. From a more general perspective, we also note the recent work of Yuen-Zhou et al. to extend the Runge-Gross theorem to open quantum systems, thereby treating the interactions between molecular electrons as well as their coupling to a class of Markovian and non-Markovian environments on equal footing \cite{AG09}.

Within our formalism, care is taken to self-consistently incorporate the polarization effects between a general quantum-mechanical molecular system and an arbitrary nanoscale metal structure supporting classical plasmon modes of excitation in the presence of the classical electromagnetic field. Generalizing and extending the original work of Gersten and Nitzan \cite{Nitzan1980}, the dynamics of the molecular adsorbate is built upon modern electronic-structure quantum-chemical theory and is simultaneously coupled to a full computational-electrodynamics description of the metallic plasmons and their associated electromagnetic fields. Symbolically, the linear response of such an interacting many-electron molecule to an external electromagnetic potential $\Phi_E$ is described by the induced electronic density 
\begin{equation}
\label{rho}
\rho^{(1)}({\bf x},t)=(1/\hbar)\tint d^3x'dt'\Pi^R({\bf x},t;{\bf x}',t')(-e)\Phi_E({\bf x}',t'),
\end{equation}
where the plasmonic polarization effects are encapsulated within the molecular polarization propagator $\Pi.$ Solution of this microscopic equation forms a large part of the following discussion, as does its self-consistent incorporation within a macroscopic-electrodynamics description of the combined molecule-metal system's electromagnetic scattering. A related, but purely theoretical development was previously reported within the context of surface-enhanced Raman scattering \cite{Masiello08a}; however, there we considered the inelastic scattering of light from a molecular target, while here we consider the elastic scattering of light from nanoscale metal particles as well as the response of the surrounding molecular environment, including their mutual coupling. Yet since much of the formalism is common, here we choose to place emphasis upon the numerical implementation and refer the interested reader to Ref. \cite{Masiello08a} for more theoretical details. As in our previous work \cite{Masiello08a}, chemical-interaction effects where the molecular wave function would extend onto the metal and vice versa are neglected.

Basic to our development is the dequantization of both the electromagnetic field $(\hat{\bf E},\hat{\bf B})\to({\bf E},{\bf B})$ and the plasmon field $\hat\Omega\to\sqrt{n}$ in their large occupation number limits, leaving only the molecular system quantum mechanical. A Born-Oppenheimer separation of electronic and nuclear degrees of freedom in the molecule facilitates the use of many-body perturbation theory to couple in the electronic polarization effects of the metal and electromagnetic field. Acting as a second polarizable body with plasmonic density $n,$ the metal, if allowed to interact with the molecular system with electronic density $\hat\rho$ through the potential $\hat V_P(t)=-e\int d^3x\hat\rho({\bf x},t)\Phi_P({\bf x},t)$ with $\Phi_P({\bf x},t)=-e\int d^3x'n({\bf x}',t)/|{\bf x}-{\bf x}'|,$ will generate a proper self-energy correction $\Sigma^\bigstar$ to the noninteracting one-body Green's function $G$ of the isolated molecule. Truncation of $\Sigma^\bigstar\approx\Sigma^\bigstar_{(2)}$ at second order in the density $\delta n=n-n_0$ induced in the metal by the molecule allows for a partial resummation of this polarization interaction through the inversion of the one-body Green's function ${\cal G}^{-1}=G^{-1}-\Sigma^\bigstar_{(2)}$ of the interacting molecule. It is through the computation of $({\cal G}^{-1})^{-1}$ that the two bodies repeatedly polarize each other, to all orders in perturbation theory, thereby allowing a self consistency to be reached between the molecule and its image as seen in the metal.

%, which, henceforth, we imagine to support a single plasmon mode of excitation at the resonant frequency $\Omega_0$ with an intrinsic line width of $\gamma_0$ and metallic density ${\cal N}_P.$ Extension to multiple resonant frequencies and widths is straighforward as $\Sigma^\bigstar_{(2)}$ is linear in ${\bm\epsilon}_P.$ Neglecting the effects of electronic correlation

We base this formalism upon an initial mean-field description of the isolated molecular system at either the Hartree-Fock (HF) or Kohn-Sham density-functional theory (DFT) level, and subsequently construct the matrix elements of $G^{-1}$ and $\Sigma^\bigstar_{(2)}$ in that basis. Adopting an isotropic Lorentz model of the metal's complex-valued polarizability ${\bm\alpha}_P$ allows for the analytic evaluation of the convolution appearing within the second-order time-ordered self energy
\begin{widetext}
\begin{equation}
\label{SE}
\begin{split}
\hbar[\Sigma^{\bigstar}_{(2)}]_{pq}(\omega)&=\sum_{rs}\langle p|-e{\bf x}|r\rangle\cdot{\bm\Lambda}\cdot{\tint}\frac{d\omega'}{2\pi i}{\bm\alpha}_{P}(\omega')G_{rs}(\omega+\omega')\cdot{\bm\Lambda}\cdot\langle s|-e{\bf x}|q\rangle\\
&\approx\frac{a_0^2}{2\omega_0 R^3}\sum_{s}\langle p|-e{\bf x}|s\rangle\cdot\Big[\frac{2}{R^3}{\bm1}+{\bm\Lambda}\Big]\cdot\langle s|-e{\bf x}|q\rangle\Big[\frac{2\omega_0}{a^2_0}{\alpha}_P(\infty)(2\rho_s^0-1)\\
&\hspace{8cm}+\frac{1-\rho_s^0}{\omega-\varepsilon^0_s/\hbar-\omega_0+i\gamma_0/2}+\frac{\rho_s^0}{\omega-\varepsilon^0_s/\hbar+\omega_0-i\gamma_0/2}\Big]
\end{split}
\end{equation} 
\end{widetext} 
with $\omega_0^2=\Omega_0^2-(\gamma_0/2)^2.$ It is expressed in terms of the noninteracting molecular orbitals $\{\phi\},$ orbital energies $\{\varepsilon^0\},$ and occupation numbers $\{\rho^0\}$ of the noninteracting molecule at its ground-state equilibrium nuclear geometry; these quantities are readily obtained from any standard electronic-structure computation. As usual, indices $i,j,k,l,\ldots$ ($a,b,c,d,\ldots$) refer to orbitals that are occupied (unoccupied) in the reference state (either mean-field or Dyson), while indices $p,q,r,s,\ldots$ are arbitrary. Unless otherwise noted, all integrals are taken over the full range of coordinates and all sums run over the complete model space. And, as in Ref. \cite{Masiello08a}, $\Lambda^{\lambda\sigma}=(3\hat R^\lambda\hat R^\sigma-\delta^{\lambda\sigma})/R^3$ with $\lambda,\sigma=x,y,z$ are matrix elements of the instantaneous near-zone dipole tensor ${\bm\Lambda}$ with ${\bf R}=R\hat{\bf R}$ the vector connecting molecule and image \cite{error}; see Fig. \ref{py}. Since it is not our intent to compute properties intrinsic to the metal, the parameters $\Omega_0,$ $\gamma_0,$ and $a_0$ representing its resonance positions, line widths, and amplitudes are chosen to fit the polarizabilities of nanoscale metallic structures within some model, e.g., Clausius-Mossotti or Gans for spheres or spheroids, or the lattice-dispersion relation \cite{draine94} for arbitrarily shaped targets, and depend upon its underlying dielectric response \cite{palik}. Extension to multiple Lorentz oscillators of the form ${\alpha}_P(\omega)\approx{\alpha}_P(\infty)-\sum_Ja_J^2/(\omega^2+i\omega\gamma_J-\Omega_J^2)$ (or even multiple Drude and/or free-electron gas oscillators) is straightforward as $\Sigma^{\bigstar}_{(2)}$ is linear in ${\bm\alpha}_{P}.$

%, thereby incorporating those image effects underlying the electromagnetic mechanism of plasmon-enhanced molecular spectroscopy. with ${\bm\epsilon}_P={\bm1}+4\pi{\cal N}_P{\bm\alpha}_P$ connecting the macroscopic and microscopic variables.  Note, that both of the commonly employed Drude and free-electron gas models occur for certain choices of these parameters.

Under the assumption that the self energy varies slowly with frequency, diagonalization of the complex-symmetric matrix ${\mathbfcal G}^{-1}(\hbar\omega)\approx{\bf G}^{-1}(\hbar\omega)-{\bm\Sigma}^\bigstar_{(2)}(\varepsilon^0)$ at each frequency $\omega$ 
\begin{equation}
\label{inv}
{\mathbfcal G}^{-1}_D(\hbar\omega)={\bf U}^{-1}(\hbar\omega){\mathbfcal G}^{-1}(\hbar\omega){\bf U}(\hbar\omega)=\hbar{\bm\omega}-{\bm\varepsilon}
\end{equation} 
results in a set of interacting and complex-valued Dyson orbitals $\{\chi\}$ making up, e.g., the columns of the right eigenvector matrix ${\bf U}$ and corresponding orbital energies $\{{\varepsilon}\}=\{{\varepsilon}^0+\hbar[{\Sigma}_{(2)}^\bigstar]_D(\varepsilon^0)\}=\{{\varepsilon}^0+\hbar{\Delta}({\varepsilon}^0)+i\hbar{\Gamma}({\varepsilon}^0)\}$ forming the matrix elements of the diagonal matrix $\hbar{\bm\omega}-{\bm\varepsilon}.$ From Eq. (\ref{inv}), the diagonal and interacting one-body molecular-electronic Green's function has the time-ordered form
\begin{equation}
\label{Ginv}
[{\cal G}^{-1}_D]_{pq}(\hbar\omega)=\delta_{pq}[(1-\rho^0_p)(\hbar\omega-\varepsilon^r_p+i\varepsilon^i_p)+\rho^0_p(\hbar\omega-\varepsilon^r_p-i\varepsilon^i_p)],
\end{equation} 
where both ${\bm\varepsilon}^r={\bm\varepsilon}^0+\hbar{\bm\Delta}({\bm\varepsilon}^0)$ and ${\bm\varepsilon}^i=\hbar{\bm\Gamma}({\bm\varepsilon}^0)$ are approximated as frequency independent, but do vary from single-particle state to single-particle state. It is the imaginary part of ${\bm\varepsilon}$ that accounts for the finite lifetime $\hbar/{\bm\varepsilon}^i$ of the electronically excited states of the molecule due to their damping by the polarization (image) interaction with the metal, while the real part is responsible for the corresponding shift of molecular energies. Construction and diagonalization of the full self-energy matrix $\hbar{\bm\Sigma}^{\bigstar}_{(2)}$ allows both of the matrices $\hbar{\bm\Delta}$ and $\hbar{\bm\Gamma}$ to be extracted.

Equipped with these Dyson orbitals and orbital energies, the density response $\delta\hat\rho=\hat\rho-\langle\hat\rho\rangle$ of the interacting molecule to an external perturbation $\hat V_E(t)=-e\int d^3x\hat\rho({\bf x},t)\Phi_E({\bf x},t)$ can be computed from the causal response kernel $\Pi^R$ known as the retarded polarization propagator. The lowest order (linear) response of the density is $\langle\delta\hat\rho({\bf x},t)\rangle_E=\rho^{(1)}({\bf x},t)+\cdots,$ where the expectation $\langle\cdots\rangle_E$ is taken in the interacting state $\langle\cdots\rangle$ additionally perturbed by $\hat V_E.$ It has the matrix elements $\rho^{(1)}_{pq}(\omega)=(1/\hbar)\sum_{rs}\Pi^R_{qprs}(\omega)(-e)\Phi^E_{rs}(\omega).$ Once $\Pi^R$ and $\hat V_E$ are known, then all observable properties related to the linear response of the system can be computed.  For example, the linear induced-dipole moment ${\bf d}^{(1)}(t)=\int dt'{\bm\alpha}_{M}(t,t')\cdot{\bf E}(t')$ is obtained from Eq. (\ref{rho}) [or its matrix elements $\rho^{(1)}_{pq}$] by multiplication by $\int d^3x(-e{\bf x})$ [or by $\sum_{pq}\langle p|-e{\bf x}|q\rangle$] in the electric-dipole interaction approximation with $\Phi_E({\bf x},t)=-{\bf x}\cdot{\bf E}(t).$ All that remains is the computation of $\Pi,$ whose matrix elements are related to the interacting one-body Green's function $\cal G$ through the convolution $\Pi_{pqrs}(\omega)=\int({d\omega'}/{2\pi i}){\cal G}_{pr}(\omega+\omega'){\cal G}_{sq}(\omega').$ We analytically perform this integral by first rotating ${\mathbfcal G}$ into its diagonal frame where it takes on the simple time-ordered form (\ref{Ginv}) and then extract its retarded component
\begin{widetext}
\begin{equation}
\label{pir}
\begin{split}
\Pi^R_{pqrs}(\omega)&\approx\sum_{p'q'r's'}U_{pp'}U_{ss'}[\Pi^R_D]_{p'q'r's'}(\omega)U^{-1}_{q'q}U^{-1}_{r'r}\\
&=\sum_{p'q'r's'}U_{pp'}U_{ss'}\delta_{p'r'}\delta_{s'q'}\frac{\rho_{q'}^0-\rho_{p'}^0}{(\varepsilon_{q'}^r-\varepsilon_{p'}^r)/\hbar+i(\varepsilon_{q'}^i+\varepsilon_{p'}^i)/\hbar+\omega}U^{-1}_{q'q}U^{-1}_{r'r}
\end{split}
\end{equation}
under the assumption that ${U}_{pq}(\omega)\approx{U}_{pq}(\varepsilon_q)\equiv{U}_{pq}.$ Rotation back to the original frame is subsequently performed numerically. By multiplication of $i\Pi^R({\bf x},t;{\bf x}',t')=\Theta(t-t')\langle[\hat\rho({\bf x},t),\hat\rho({\bf x}',t')]\rangle$ with $\int d^3xd^3x'(-e{\bf x})(-e{\bf x}')$ or contraction of $\rho^{(1)}_{pq}$ with $\langle p|-e{\bf x}|q\rangle,$ the one-body polarizability ${\bm\alpha}_M$ of the molecule including its polarization (image) interaction with a nearby plasmon supporting metal is found to be
\begin{equation}
\label{alpha}
\begin{split}
\alpha_M^{\lambda\sigma}(\omega)&=-\frac{1}{\hbar}\sum_{pqrs}\langle q|-e{x}^\lambda|p\rangle\Pi^R_{pqrs}(\omega)\langle r|-e{x}^\sigma|s\rangle\\
&=\sum_{ia}\Big[\frac{(a|d^\lambda|i)(i|d^\sigma|a)}{(\varepsilon_a^r-\varepsilon_i^r)+i(\varepsilon_a^i+\varepsilon_i^i)+\hbar\omega}+\frac{(a|d^\sigma|i)(i|d^\lambda|a)}{(\varepsilon_a^r-\varepsilon_i^r)-i(\varepsilon_a^i+\varepsilon_i^i)-\hbar\omega}\Big]
%\alpha_M^{\lambda\sigma}(\omega)&=-\frac{1}{\hbar}\sum_{pqrs}\langle q|-e{x}^\lambda|p\rangle\Pi^R_{pqrs}(\omega)\langle r|-e{x}^\sigma|s\rangle\\
%&=\sum_{ia}\Big[\frac{(a|d^\lambda|i)(i|d^\sigma|a)}{(\varepsilon_a^r-\varepsilon_i^r)+i(\varepsilon_a^i+\varepsilon_i^i)+\hbar\omega}+\frac{(a|d^\sigma|i)(i|d^\lambda|a)}{(\varepsilon_a^r-\varepsilon_i^r)-i(\varepsilon_a^i+\varepsilon_i^i)-\hbar\omega}\Big]
\end{split}
\end{equation}
\end{widetext}
with one-body dipole-matrix elements $(p|d^\lambda|q)=\sum_{rs}U^{-1}_{ps}\langle s|-e{x}^\lambda|r\rangle U_{rq}$ rotated into the Dyson basis denoted by the symbol $|\,).$ This expression generalizes the standard one-body approximation to the Kramers-Heisenberg polarizability and further extends the work of Norman et al. \cite{Oddershede01} and Jensen et al. \cite{Jensen05}, where the infinitesimals $0^+$ appearing in the denominators of the molecular polarizability are replaced by an empirical parameter $\hbar\Gamma_{\textrm{exp}}$ that mimics the molecule-metal interaction and defines the polarizability on resonance, to explicitly and rigorously include the resonance widths $\varepsilon_a^i+\varepsilon_i^i.$ It is, of course, important to note that Eq. (\ref{alpha}) reduces to the usual one-body molecular polarizability of an isolated molecule in the limit where ${\bm\varepsilon}^r={\bm\varepsilon}^0$ and ${\bm\varepsilon}^i=0^+,$ i.e., when $\hat V_P\to0.$ In that case ${\bf U}\to{\bf 1}$ and $(p|d^\lambda|q)\to\langle p|-e{x}^\lambda|q\rangle.$

To go beyond the mean-field approximation of the molecular electron-electron interactions in the calculation of the ${\bm\alpha}_M,$ perturbation theory may be additionally performed on the time-ordered polarization propagator $\Pi.$ Namely, the two-body Coulomb potential $\hat V_M(t)=\int d^3xd^3x'\hat\rho({\bf x},t)V_M({\bf x},{\bf x}')\hat\rho({\bf x}',t)=(e^2/2)\int d^3xd^3x'\hat\rho({\bf x},t)\hat\rho({\bf x}',t)/|{\bf x}-{\bf x}'|$ generates a proper self-energy correction $K^\bigstar$ to the interacting $\Pi$ of the molecule-metal system. Truncation of $K^\bigstar\approx K^\bigstar_{(1)}$ at first order in perturbation theory allows for a partial resummation of the effects of electron-electron interaction at the level of the random-phase approximation (RPA), through the inversion of the RPA polarization propagator $\Pi_{\textrm{RPA}}^{-1}=\Pi^{-1}-K^\bigstar_{(1)};$ see, e.g., Ref. \cite{FW}. It is at the poles of $\Pi_{\textrm{RPA}}$ or the zero eigenvalues of $\Pi_{\textrm{RPA}}^{-1}$ that the low-amplitude collective excitations of the molecular-electronic system are located. These zeroes can be found by solving the associated pseudoeigenvalue problem
\begin{equation}
\label{pirpa}
[{\bm\Pi}^{-1}_{\textrm{RPA}}]_D(\omega){\bm\rho}^{(1)}(\omega)=[{\bm\Pi}_D^{-1}(\omega)-{\bf K}^\bigstar_{D(1)}(\omega)]{\bm\rho}^{(1)}(\omega)={\bf 0}
\end{equation}
in the Dyson basis, where $[{\Pi}_{\textrm{RPA}}^{-1}]_{pqrs}=\sum_{p'q'r's'}U_{pp'}U_{ss'}[{\Pi}^{-1}_{\textrm{RPA}}]^D_{p'q'r's'}U^{-1}_{q'q}U^{-1}_{r'r}$ and $\hbar[K^\bigstar_{D(1)}]_{pqrs}=-(ps|V_M|rq)+(sp|V_M|rq).$ Eqs. (\ref{pirpa}) may be decomposed into $(i,a)$ and $(a,i)$ components
\begin{widetext}
\begin{equation}
\label{gen}
\begin{split}
[(\varepsilon_{a}^r-\varepsilon_{i}^r)-i(\varepsilon_{a}^i+\varepsilon_{i}^i)-\hbar\omega]\rho^{(1)}_{ia}(\omega)+\sum_{jb}&\big\{[(ja|V_M|bi)-(aj|V_M|bi)]\rho^{(1)}_{jb}(\omega)\\
&+[(ba|V_M|ji)-(ab|V_M|ji)]\rho^{(1)}_{bj}(\omega)\big\}=0\\
[(\varepsilon_{a}^r-\varepsilon_{i}^r)-i(\varepsilon_{a}^i+\varepsilon_{i}^i)+\hbar\omega]\rho^{(1)}_{ai}(\omega)+\sum_{jb}&\big\{[(ja|V_M|bi)-(aj|V_M|bi)]^*\rho^{(1)}_{bj}(\omega)\\
&+[(ba|V_M|ji)-(ab|V_M|ji)]^*\rho^{(1)}_{jb}(\omega)\big\}=0,
\end{split}
\end{equation}
\end{widetext}
where they generalize the well-known RPA equations to complex-valued orbitals and orbital energies. In the noninteracting limit, i.e., when $\hat V_P\to0,$ the standard RPA equations are recovered with $A_{iajb}=\delta_{ij}\delta_{ba}(\varepsilon_{a}^0-\varepsilon_{i}^0)+\langle ja|V_M|bi\rangle-\langle aj|V_M|bi\rangle$ and $B_{iajb}=\langle ba|V_M|ji\rangle-\langle ab|V_M|ji\rangle.$ Solution of these generalized RPA equations (\ref{gen}) followed by a rotation back to the original basis results in new eigenenergies that account for all possible single-electron excitations and single-electron de-excitations from the interacting molecular reference system in the Condon approximation; setting $B=0$ would be equivalent to performing a configuration interaction with all possible single excitations from the mean-field reference, also called the Tamm-Dancoff approximation. The resulting polarizability $-\hbar{\bm\alpha}^M_{\textrm{RPA}}(\omega)=\sum_{pqrs}\langle q|-e{\bf x}|p\rangle([{\Pi}_{\textrm{RPA}}^{R}]^{-1})^{-1}_{pqrs}(\omega)\langle r|-e{\bf x}|s\rangle$ built from these energies would improve upon that in Eq. (\ref{alpha}) to include this limited class of electronic excitations.

%!!!!!!!!!!!!!!!!!!!!!!!!!!!!!!!!!!!!!!!!!!!!!!!!!!!!!!!!!!!
%!!!!!!!!!!!!!!!!!!!!!!!!!!!!!!!!!!!!!!!!!!!!!!!!!!!!!!!!!!!
\begin{figure*}[t]
\begin{center}
\rotatebox{0}{\resizebox{!}{8.5cm}{\includegraphics{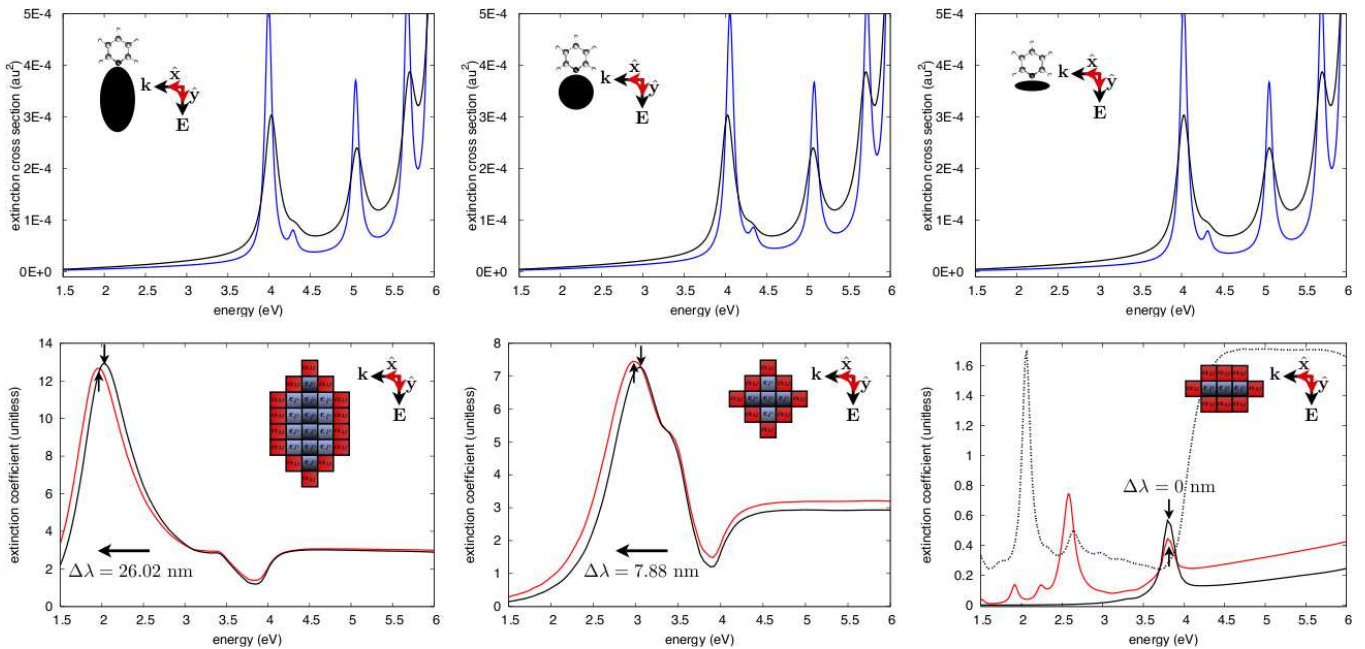}}}
\vspace{-0.45cm}
\caption{\label{alphafig} Averaged extinction cross sections $\sigma^M_{\textrm{ext}}=4\pi k\textrm{Im}[\alpha_M^{xx}+\alpha_M^{yy}+\alpha_M^{zz}]/3$ (upper) and extinction coefficients $\sigma_{\textrm{ext}}/$(cross-sectional area) (lower) of the coupled pyridine-silver nanospheroid system as a function of excitation energy and plasmon-resonance position. Explicit treatment of the repeated polarization interaction between a single pyridine molecule and its image in a nearby metal spheroidal particle reveals that the pyridine polarizability and extinction (upper, blue) are sensitive to nanoparticle shape, LSPR position and width, and relative molecule-metal separation and orientation, here chosen so that molecule and image are collinear with each other ($\varphi=0$) and with the field polarization direction, and separated by $R=1.6$ nm. For comparison, we also compute the pyridine extinction cross section based upon an empirical and constant parametrization of the molecule-metal interaction $\Sigma^\bigstar\approx i\Gamma_{\textrm{exp}}$ (upper, black) that is consistent with the work in Ref. \cite{Jensen05}. Progression from a pyridine-coated oblate spheroid (right) with radii (50,1,50) nm, to sphere (middle) with radii (50,50,50) nm, to prolate spheroid (left) with radii (50,100,50) nm allows for the $y$-axis plasmon resonance to be tuned from nearly resonant with to red-detuned from the lowest-lying molecular resonance of the interacting molecule. Extinction coefficients of the coupled system in the monolayer-coverage limit (lower, red) and bare metal (lower, black) are shown with the excitation field polarized along the $y$ axis as well as polarized along the $z$ axis in the oblate case for the bare metal only (lower right, dotted black). Insets not drawn to scale.}
\end{center}
\end{figure*}
%!!!!!!!!!!!!!!!!!!!!!!!!!!!!!!!!!!!!!!!!!!!!!!!!!!!!!!!!!!!
%!!!!!!!!!!!!!!!!!!!!!!!!!!!!!!!!!!!!!!!!!!!!!!!!!!!!!!!!!!! 

We numerically implement these molecular interactions with the metal and the external field due to the perturbations $\hat V_P$ and $\hat V_E$ as a post-processing step within a modified version of the Q-Chem quantum-chemistry software package \cite{qchem}, but do not treat any RPA-level electronic excitations induced by $\hat V_M$ at this time. These former interactions are built upon an initial mean-field calculation of the molecule's electronic ground state, which, for the examples presented in the following, will be DFT. Dyson orbitals and corresponding orbital energies are not brought to self consistency in the sense that the self energy (\ref{SE}) and interacting one-body Green's function (\ref{inv}) are computed directly and only once from the noninteracting Kohn-Sham orbitals and orbital energies $\{\phi\}$ and $\{\varepsilon^0\}.$ Continuum-electromagnetic scattering properties of the combined metal-molecule system are subsequently computed through a version of the discrete-dipole approximation (DDA) \cite{draine94} that has been modified to directly accept the microscopic and isotropic molecular polarizability as input rather than the molecular system's effective macroscopic index of refraction or dielectric function.

%self consistency between the molecular Green's function and its underlying orbitals and orbital energies is saved for future work and we do not expect its ommission here to qualitatively change our results. 

As a first application of our formalism, we compute the linear response and scattering of pyridine interacting with a nearby nanoscale silver spheroidal particle; see Fig. \ref{alphafig}. By deforming the spheroid's aspect ratio from prolate (left panel) with radii (50,100,50) nm, to sphere (middle panel) with radii (50,50,50) nm, to oblate (right panel) with radii (50,1,50) nm, the particle's $y$-axis plasmon resonance may be tuned across a portion of the electronic spectrum of the pyridine-nanospheroid system, thereby allowing us to systematically explore their interaction. Sensitivity of the LSPR position to the surrounding environment's refractive index in the monolayer-coverage limit of a variety of electronically resonant molecules on silver nanotriangles has recently been observed \cite{RVD06,Zhao07} and serves to motivate our first-principles calculations.

%An analogous experiment \cite{Zhao07} has already demonstrated the sensitivity of the plasmon resonance position to the surrounding environment's refractive index in the monolayer-coverage limit of rhodamine 6G on silver nanotriangles, and serves to motivate our first-principles calculations.

Our computations involve two steps: first, the microscopic quantum-mechanical response of a single pyridine molecule is computed as it interacts with its image in a nearby silver nanoscale spheroidal particle, described analytically by either Clausius-Mossotti or Gans models for ${\bm\alpha}_P({\bm\epsilon}_P)$; see Fig. 2, upper panels. Second, the resulting isotropic pyridine polarizability is incorporated into a DDA description of the macroscopic electromagnetic scattering of a monolayer-covered silver spheroid, where the monolayer and metal are discretized into a collection of polarizable points with polarizabilities defined from step one for the molecule and by the lattice-dispersion relation \cite{draine94} for the metal; see Fig. 2, lower panels. In both steps the underlying dielectric data for silver is taken from Ref. \cite{palik}.

To further elaborate, first, we fit multiple Lorentz oscillators to ${\bm\alpha}_P,$ which together with a DFT-VWN ground-state electronic-structure calculation in a 6-31g* basis, define the self energy $\Sigma^{\bigstar}_{(2)}$ (\ref{SE}), one-body Green's function ${\cal G}$ and polarization propagator $\Pi^R$ (\ref{pir}), and polarizability ${\bm\alpha}_M$ (\ref{alpha}) of the interacting molecule. From the imaginary component of ${\bm\alpha}_M,$ we display the averaged extinction cross section $\sigma^M_{\textrm{ext}}$ of a single pyridine molecule as a function of energy and spheroid aspect ratio in Fig. \ref{alphafig} (upper, blue), together with that computed (upper, black) using the approximations of Refs. \cite{Oddershede01,Jensen05}; the latter chosen to empirically parametrize the molecule-metal interaction $\Sigma^\bigstar\approx i\Gamma_{\textrm{exp}}$ directly by a line width of $\hbar\Gamma_{\textrm{exp}}=0.1$ eV that is independent of molecular state, energy, relative molecule-metal orientation and separation, and plasmonic structure. To compare both approaches, we orient pyridine's dipole moment to be normal to the metallic surface (a choice of $\varphi=0$ in the representation shown in Fig. \ref{py}), along the field-polarization axis, and separated from its image by $R=1.6$ nm; this distance is chosen because it falls in the middle of a regime, from $\sim1-3$ nm, where image effects seem to be important, i.e., are comparable in magnitude to the homogeneous broadening, while smaller separations reveal a (expected) breakdown of our perturbative approach. We also adopt a reasonable and constant value for the homogeneous line width of pyridine to be $\hbar\gamma_H=0.05$ eV based on analogy with estimates for other molecules \cite{Masiello09,Schatz05,Myers88}. Since our model only describes the modification of the molecule's dynamics due to its interaction with the plasmon and does not include any intrinsic- or homogeneous-broadening mechanisms it is necessary to add $\hbar\gamma_H$ before comparing with experiment; however, we assume that such mechanisms are completely decoupled from the relevant plasmon enhancement, thereby justifying such a separation. In Fig. \ref{alphafig}, it is important to note that the shifting and broadening of molecular-excited states due to their interaction with the metal are dependent not only upon the LSPR position of the metal but also upon its strength; e.g., the lowest-lying pyridine excited state is nearly resonant with the oblate spheroid's $y$-axis dipole-plasmon resonance, yet its interaction with the oblate spheroid is not as strong as with the red-detuned sphere or prolate spheroid because of the negligible strength of the oblate spheroid's LSPR; note the change in magnitude of the extinction coefficient at the LSPR maximum of the prolate spheroid, sphere, and oblate spheroid in the lower three panels of Fig. \ref{alphafig}. Maximizing the two, oscillator strength and spectral overlap, would lead to the greatest interaction effects modifying the molecular polarizability.

Second, we add a pyridine monolayer to a DDA-level continuum-electrodynamics calculation of the bare silver spheroid, where the monolayer and metal are built from a collection of polarizable points with isotropic polarizabilities ${\bm\alpha}_M$ computed from step one and ${\bm\alpha}_P({\bm\epsilon}_P)$ computed from the lattice dispersion relation \cite{draine94}; each DDA point occupies a cubic volume of $(2\textrm{ nm})^3$ with other choices of nearby dipole spacing yielding similar results. Here we point out that while ${\bm\alpha}_M$ accounts for the repeated polarization interaction between molecule and metal, it does not include any direct quantum-mechanical molecule-molecule couplings; however, it would be interesting to explore the effect of different molecular aggregation schemes \cite{Zhao07} and how they interact with the metal with our methodology in the future. Numerical solution of this combined DDA accounts for the self-consistent electric-dipole interaction between all of the metallic and molecular dipoles among themselves and, additionally, with each other. The resulting extinction coefficients $\sigma_{\textrm{ext}}/$(cross-sectional area) of the combined and coupled system (lower, red) as well as those of the bare silver spheroid (lower, black) are displayed in Fig. \ref{alphafig} as a function of excitation energy and spheroid aspect ratio. The progression from no shift ($\Delta\lambda=0$ nm, oblate) to a red shift (sphere, $\Delta\lambda=7.88$ nm; prolate, $\Delta\lambda=26.02$ nm) of the combined system's LSPR wavelength $\lambda$ with respect to that of the bare metal, as the plasmonic and lowest-lying molecular resonances are detuned, is consistent with the observations reported in Refs. \cite{RVD06,Zhao07}. Also note that while the pyridine monolayer does not affect the position of the oblate spheroid's $y$-axis dipole-plasmon resonance, it does mediate the coupling of incident light to the previously dark $z$-axis plasmon resonance; in particular, the feature appearing at 2.6 eV (lower right, red) should be compared to the new plasmon resonance at 2.7 eV of the bare oblate spheroid excited with $z$-polarized light (lower right, dotted black).

%Another interesting feature evident in the cases of the prolate spheroid's and sphere's extinction spectra (lower left and lower center) is the pronounced dip just below 4 eV. We understand this to be due to the interaction of the lowest lying molecular-electronic excited state of pyridine with the blue edge of the dipole plasmon's near continuum resonance, where the molecular excitation steals oscillator strength from the LSPR; such a feature would also be called a Fano antiresonance.

%!!!!!!!!!!!!!!!!!!!!!!!!!!!!!!!!!!!!!!!!!!!!!!!!!!!!!!!!!!!
%!!!!!!!!!!!!!!!!!!!!!!!!!!!!!!!!!!!!!!!!!!!!!!!!!!!!!!!!!!!
\begin{figure}[t]
\begin{center}
\rotatebox{0}{\resizebox{!}{6cm}{\includegraphics{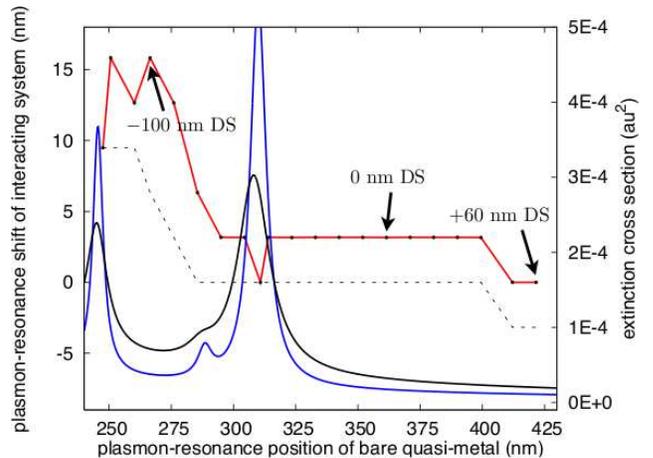}}}
%\vspace{-0.45cm}
\caption{\label{shift} Plasmon-resonance shift of a pyridine-coated 20-nm radius quasi-silver nanosphere as a function of the bare sphere's tunable plasmon-resonance position (black bullets, connected in red). Variation, in increments of 10 nm, of the sphere's underlying dielectric response (from -130 nm DS to +60 nm DS) with respect to that of silver (0 nm DS taken from Ref. \cite{palik}) allows us to tune the bare quasi-metal's LSPR across several low-lying molecular-electronic excitations. For comparison the PRS of a 20-nm radius quasi-silver nanosphere coated with a nonresonant monolayer with index of refraction $n=1.5$ is displayed as the dashed black curve. Corresponding to a 0 nm DS, the computed extinction spectrum of a single pyridine molecule interacting with its image in the silver nanosphere (blue) is superimposed for reference along with that computed from an empirical treatment of the pyridine-metal interaction (black) consistent with Ref. \cite{Jensen05}.}
\end{center}
\end{figure}
%!!!!!!!!!!!!!!!!!!!!!!!!!!!!!!!!!!!!!!!!!!!!!!!!!!!!!!!!!!!
%!!!!!!!!!!!!!!!!!!!!!!!!!!!!!!!!!!!!!!!!!!!!!!!!!!!!!!!!!!! 

As a further test of our formalism we examine the fictitious interaction of pyridine with a 20-nm radius quasi-silver sphere, whose underlying dielectric function \cite{palik} is shifted, by hand, to allow its dipole-plasmon resonance to be tunable across several low-lying molecular-electronic excitations. (This is normally not possible for pyridine-silver nanoparticle system.) By varying this dielectric shift (DS), each bullet in Fig. \ref{shift} (connected in red) represents a separate computation of the extinction spectrum and resulting plasmon-resonance shift (PRS) of the combined molecule-quasi-metal system with respect to that of the bare quasi-metal; 0 nm DS corresponds to the real pyridine-silver system with an LSPR wavelength of 370 nm. For comparison, the PRS of a 20-nm radius quasi-silver nanosphere coated with a uniform and nonresonant layer with index of refraction $n=1.5$ is shown as the dashed black curve. Superimposed on the plot is the extinction cross section of a single pyridine molecule (blue) interacting with its image, located $R=1.6$ nm away and with dipole moment aligned with the field-polarization direction (a choice of $\varphi=0$ in the representation shown in Fig. \ref{py}), in the silver sphere (for the case having 0 nm DS), as well as that computed from an empirical parametrization of the molecule-metal interaction (black) consistent with Ref. \cite{Jensen05}. (These curves are analogous to the blue and black curves in the upper panels in Fig. \ref{alphafig}.) Our computations reveal that the PRS increases as the DS becomes more negative. This is the expected result for a nonresonant scatterer interacting with a metal particle, as the shift of the plasmon resonance should become larger as the particle becomes more polarizable, and in the present application this corresponds to going to the blue. However, we also notice dips in the PRS at $\sim$315 nm and $\sim$245 nm, which coincide with molecular resonances in pyridine (blue curve). The dips are understood to arise because the localized but negative (anomalous) polarizability of pyridine destructively interferes with the positive (normal) portion of that of the particle as the LSPR passes through resonance \cite{RVD06,Zhao07}. This behavior is related to behavior found in the experiments in Refs. \cite{RVD06,Zhao07}, which show a pronounced dip in the PRS as the plasmon is tuned through the molecular resonance. In this case, the dips are smaller due to the weakness of the pyridine transitions, but the physical effect is the same.

In conclusion, we present the first numerical implementation of our general many-body description of a variety of plasmon-enhanced linear spectroscopies. We briefly review our theoretical formalism in order to motivate and detail our computational efforts in linking together both Q-Chem \cite{qchem} and the DDA \cite{draine94} to self-consistently and practically describe the linear-response and scattering properties of an arbitrary quantum-mechanical molecule interacting with a nearby classical nanoscale metal in the presence of an external continuum electromagnetic field. As a preliminary test of our methodology we compute the linear response and scattering of the pyridine-silver nanospheroid system as a function of excitation energy and spheroid aspect ratio.  Variation of the latter allows us to tune the nanoparticle's LSPR position from nearly resonant with to red-detuned from the molecule's lowest-lying electronic excitation; the positions and widths of such are here computed from first principles. Interactions between these resonances are highlighted and qualitatively discussed in connection to recent experimental observations of similar systems \cite{RVD06,Zhao07}. In addition, we examine the fictitious interaction between pyridine and a quasi-silver nanosphere with tunable LSPR position in order to further explore the coupling of the plasmon to multiple electronic excitations in the molecule. Quantitative and rigorous modeling of a variety of plasmon-enhanced linear elastic and inelastic spectroscopies is now enabled with our approach and will serve to motivate our future investigations.

\begin{acknowledgments}
The authors gratefully acknowledge financial support from DOE grant DE-SC0001785. Further, D.J.M. has benefited from many helpful conversations with Dr. Joseph E. Subotnik on the numerical implementation of the interacting molecular polarizability within Q-Chem, and with Drs. David S. Ginger and Markus B. Raschke on the experimental interpretation of the presented data.
\end{acknowledgments}

\bibliography{jila,thesis,mal,ref}

\begin{thebibliography}{28}
\bibitem[{err()}]{error}
\bibinfo{note}{Eq. (58) in Ref. \cite{Masiello08a} should be corrected to
  reflect that
  $R^3\Lambda^{\lambda\eta}\Lambda^{\eta\sigma}=(2/R^3)\delta^{\lambda\sigma}+%
\Lambda^{\lambda\sigma}$.}
\end{thebibliography}

\end{document}